# Are ChatGPT and Other Similar Systems the Modern Lernaean Hydras of AI?[*]

Dimitrios Ioannidis, Esq.,[**] Dr. Jeremy Kepner,[***] Dr. Andrew Bowne, Lt. Col., USAF,[****] Harriet S. Bryant[*****]

---

[**] Dimitrios Ioannidis, Esq. is a partner in Roach, Ioannidis & Megaloudis, LLC based in Boston, founder of the Innovation Moot (www.innovationmoot.com), and a co-founder of OsmoCosm, a non-profit think tank that supports emerging technologies in olfaction and promotes the ethical collection and use of olfactory data. (www.osmocosm.org). He is also a core team member of www.Mycohab.com, a company that uses mycelium technology from lab to market to develop an innovative way of building affordable houses and in the process, creating jobs, contributing to food security, and rehabilitating the local ecosystem in Namibia. He recently authored a law review article on whether AI will replace arbitrators under the Federal Arbitration Act. *See* Dimitrios Ioannidis, *Will Artificial Intelligence Replace Arbitrators Under the Federal Arbitration Act?*, 28 RICH. J.L. & TECH. 505 (2022). He received a JD from Boston University School of Law and a BA in economics and political science from Boston University. *See Dimitrios Ioannidis*, LINKEDIN, https://www.linkedin.com/in/dimitrios-ioannidis-4783258/ [perma.cc/JU2W-C474].

[***] Dr. Jeremy Kepner is head and founder of the MIT Lincoln Laboratory Supercomputing Center (LLSC), and also a founder of the MIT-Air Force AI Accelerator. Additional appointments include the MIT Mathematics Department and MIT Connection Science. Dr. Kepner received a PhD in Astrophysics from Princeton University and a BA in Astrophysics from Pomona College in 1991. Dr. Kepner's research is described in hundreds of peer-reviewed papers spanning computing, science, and mathematics. *See Dr. Jeremy Kepner (Supercomputing Center Head & Founder)*, http://www.mit.edu/~kepner [perma.cc/X8ER-R47A].

*The rise of Generative Artificial Intelligence systems ("AI systems") has created unprecedented social engagement. AI code generation systems provide responses (output) to questions or requests by accessing the vast library of open-source code created by developers over the past few decades. However, they do so by allegedly stealing the open-source code stored in virtual libraries, known as repositories. This Article focuses on how this happens and whether there is a solution that protects innovation and avoids years of litigation. We also touch upon the array of issues raised by the relationship between AI and copyright. Looking ahead, we propose the following: (a) immediate changes to the licenses for open-source code created by developers that will limit access and/or use of any open-source code to humans only; (b) we suggest revisions to the Massachusetts Institute of Technology ("MIT") license so that AI systems are required to procure appropriate licenses from open-source code developers, which we believe will harmonize standards and build social consensus for the benefit of all of humanity, rather than promote profit-driven centers of innovation; (c) we call for urgent legislative action to protect the future of AI systems while also promoting innovation; and (d) we propose a shift in the burden of proof to AI systems in obfuscation cases.*

**** Dr. Andrew Bowne is an active-duty Air Force Judge Advocate, currently assigned as the Chief Legal Counsel of the Department of the Air Force – MIT Artificial Intelligence Accelerator. Dr. Bowne is the author of dozens of articles, book chapters, and policies focused on the intersection of law and technology, including intellectual property, government contracts, and ethics. He received a PhD from the University of Adelaide, a LLM with a specialization in Contract and Fiscal Law from the Judge Advocate General's School, a JD from the George Washington University Law School, and a BA in political science from Pepperdine University. *See Andrew Bowne*, LINKEDIN, https://www.linkedin.com/in/andrew-bowne-8b9918b0/ [perma.cc/AB6T-AU4J].
***** Harriet S. Bryant is a third-year student at Suffolk University Law School and Chief Comment Editor of the Suffolk Transnational Law Review. She took on the initial task of drafting parts of this article.

INTRODUCTION

The rise of Generative Artificial Intelligence systems ("AI systems") parallels the Greek myth of Pandora who was overwhelmed with curiosity and opened the Box, "[r]eleasing curses upon mankind."[1] However, Pandora's Box is not solely about evil or curses as the artifact-looking Box included Elpis, the personified spirit of Hope,[2] which is a clear reminder that a lot of good can come out of the development of AI systems. To add to the analogies here, AI systems can be thought of as "the monster plant Audrey II in *Little Shop of Horrors*, constantly crying out 'Feed me!'"[3] Why? Because ChatGPT and other AI systems provide a natural language response (output) to questions or requests by accessing the vast library of open-source code created by developers over decades.

---

[1] *See Pandora's Box*, WIKIPEDIA, https://en.wikipedia.org/wiki/Pandora%27s_box [https://perma.cc/LZY4-BWKH] (last visited Sept. 5, 2023).

But what happens if there is a lapse in this constant frenzy of "feeding" the AI systems with updated data? Can they continue to operate if there is an interruption in their supply chain of "food" (data) or will they "starve" to the point of extinction? The answers to these questions are exceedingly perplexing. Currently, open-source code and the data generated are uploaded by their developers and stored in virtual libraries (known as repositories) for the benefit of all humanity, with some strings attached. These "strings" formed the basis of a class action complaint filed by a group of John Doe plaintiffs against GitHub, Inc. ("GitHub"), Microsoft Corporation ("Microsoft"), and OpenAI, Inc. ("OpenAI") in the Northern District of California in November 2022 (hereinafter "*J. Doe v. GitHub, Inc.*").[4] In general, the *J. Doe v. GitHub, Inc.* plaintiffs contend that the defendants (through their AI systems) committed intellectual property theft by pilfering data from developers of open-source code.[5]

---

[2] *See* Rittika Dhar, *Pandora's Box: The Myth Behind the Popular Idiom*, HISTORY COOP. (Aug. 17, 2022), https://historycooperative.org/pandoras-box/ [perma.cc/D9B4-KCPB].

[3] *See* Preston Gralla, *This Lawsuit Against Microsoft Could Change the Future of AI*, COMPUTERWORLD (Jan. 10, 2023), https://www.computerworld.com/article/3684734/this-lawsuit-against-microsoft-could-change-the-future-of-ai.html [perma.cc/6QCE-HYNG] (analogizing AI systems).

[4] *See* Complaint at 1, J. Doe v. GitHub, Inc., No. 44:22-cv-06823 (N.D. Cal. Nov. 3, 2022) [hereinafter Compl.] (stating parties in action). There are also several other cases that have been filed against generative AI companies. *See, e.g.*, Chabon v. OpenAI, Inc., No. 3:23-cv-04625 (N.D. Cal. filed Sept. 8, 2023); J.L. v. Alphabet, Inc., No. 3:23-cv-03440-LB (N.D. Cal. filed July 11, 2023); Silverman v. OpenAI, Inc., No. 3:23-cv-03416 (N.D. Cal. filed July 7, 2023); Kadrey v. Meta Platforms, Inc.*,* No. 3:23-cv-03417 (N.D. Cal. filed July 7, 2023); Tremblay v. OpenAI, Inc., No. 3:23-cv-03223 (N.D. Cal. filed June 28, 2023); P.M. v. OpenAI LP*,* No. 3:23-cv-03199 (N.D. Cal. dismissed Sept. 15, 2023); Walters v. OpenAI, L.L.C., No. 123-cv-03122 (N.D. Ga. Filed July 14, 2023); Young v. NeoCortext, Inc.*,* No. 2:23-cv-02496 (C.D. Cal. filed Apr. 3, 2023); Flora v. Prisma Labs, Inc., No. 323-cv-00680 (N.D. Cal. filed Feb. 15, 2023); Getty Images v. Stability AI, Inc.*,* No. 1:2023-cv-00135 (D. Del. filed Feb. 3, 2023); Andersen v. Stability AI Ltd. No. 3:23-cv-00201 (N.D. Cal. filed Jan. 13, 2023).

This Article endeavors to address how this theft happens and whether there is a solution, short of years of litigation, which can protect innovation while recognizing the rapid expansion of AI systems and the conflicting interests currently at play.[6] We also peripherally touch upon an array of issues raised by the relationship between AI and copyright, including the conflicting enforcement of licensing agreements and Terms of Service governing open-source code repositories, whether the fair use defense could dictate the scope of open-source code usage or AI-generated works, and the role (or perhaps the necessity) of achieving social consensus through human supervision of the future of AI.[7] By looking ahead, we propose the following.

First, we call for immediate changes to the licenses for open-source code created by developers and propose new language which will limit access and/or use of any open-source code *to humans only*.[8] This is not only based on the premise that humans must always be in control of the input of AI systems, but also on the idea that we must shape the fundamental principles for such emerging technologies now—not years later, when AI systems may be exceedingly difficult to regulate and monitor. Next, we suggest revisions to the Massachusetts Institute of Technology ("MIT") open-source code license so that ChatGPT and other AI systems procure the appropriate licenses from code developers. We believe this will harmonize standards and build social consensus for the benefit of all of humanity rather than the profit-driven centers of innovation. Further, we make an urgent call for legislative action to protect the future of AI systems while also promoting innovation. The action must include safeguards to eliminate the raiding of open-source code and data, along with the risk of private or governmental acts of extremism.[9] Finally, we propose that the burden of proof shift from plaintiffs to AI systems in obfuscation cases, which will make it more difficult for owners of AI systems to use "statistical tracing" or similar arguments to defeat the claims.[10]

---

[5]    *See* Compl., *supra* note 4 (stating causes of action).
[6]    *See Artificial Intelligence: Stop to ChatGPT by the Italian SA Personal Data is Collected Unlawfully, No Age Verification System Is in Place for Children*, GARANTE PRIV. (Mar. 31, 2023), https://www.garanteprivacy.it/home/docweb/-/docweb-display/docweb/9870847#english [https://perma.cc/B5PD-NBAR] (reporting Italian Data Protection Authority's temporary prohibition). The Italian SA has instituted a temporary prohibition on ChatGPT, an AI platform developed and managed by OpenAI, following a reported data breach of ChatGPT's users' conversations and payment information. *Id.* The Italian SA also alleges the system is collecting large swathes of personal data to "train" the algorithms without any legal basis. *Id.*
[7]    *See* discussion *infra* Parts II A, C, F.
[8]    *See infra* Exhibits 2–3.

## A. The "Rise" of the Open-Source Code[11]

Data, known as "open-source code," is stored by its developers in vast public repositories, such as GitHub, where it is subsequently pulled by AI software to generate output with "human-like skill."[12] There are conditions attached to using the open-source code. The code is often copyrighted and only accessible to users if they either agree to certain licensing terms or attribute the open-source code to the named developer (usually accompanied by a copyright notice).[13] MIT first developed such a license in the 1980s and it has since grown to be the most popular open-source license agreement, accounting for 27% of licenses, because "it's short and to the point."[14] Besides the MIT license, there is also an emergence of free license agreements circulating the internet which provide developers with the means to both protect and promote their source code when it is in other developers' hands.[15]

---

[9] *See DILEMA 2023 Conference: Call for Abstracts*, ASSER INST. (Apr. 17, 2023), https://www.asser.nl/dilema/news-and-updates/dilema-2023-conference-call-for-abstracts/ [perma.cc/4A4W-FWCQ] (organizing military AI conference). The DILEMA Project's conference centers around the complex and interdisciplinary issues raised by military applications of artificial intelligence. *Id.*

[10] *See* discussion *infra* Parts II, C, D, E and Part III; *see also* Int'l Bd. of Teamsters v. United States, 431 U.S. 324, 359 n.45 (1977) ("Presumptions shifting the burden of proof are often created to reflect judicial evaluations of probabilities and to conform with a party's superior access to the proof.").

[11] *See* TERMINATOR 3: RISE OF THE MACHINES (Warner Bros. Pictures 2003); *see also* Larry Wasserman, *Rise of the Machines*, CARNEGIE MELLON U.: STAT. & DATA SCI., https://www.stat.cmu.edu/~larry/Wasserman.pdf [https://perma.cc/SJC7-924W] (last visited Aug. 5, 2023) (contemplating the evolution of machine learning).

[12] *See* Gralla, *supra* note 3 (describing how AI is trained on open-source code repositories); *see also What is Open Source?*, OPEN SOURCE, https://opensource.com/resources/what-open-source [https://perma.cc/4RHR-HRFL] (last visited Aug. 5, 2023) ("Source code is the part of software that most computer users don't ever see; it's the code computer programmers can manipulate to change how a piece of software—a 'program' or 'application'—works."); Jason Wise, *How Much Data is Generated Every Day in 2023?*, EARTHWEB (Feb. 21, 2023), https://earthweb.com/how-much-data-is-created-every-day/ [https://perma.cc/N9HW-3ZSL] (showing that in 2023, 3.5 billion quintillion bytes of data is created every day which float around on digital "clouds"). For context, 1 GB of data creates around 350,000 emails. *Id.*

GitHub was launched in 2008 as a forum to "support open-source development."[16] By providing software developers a platform to publish licensed materials, which required some form of attribution and copyright notice of the developer, GitHub set itself apart as the de facto software sharing platform.[17] To encourage

---

[13] *See What is Open Source?*, *supra* note 12 (describing attribution protocols). Think of repositories as a multi-level parking garage. The owner of each vehicle enters a checkpoint, where at the click of a button the barrier lifts and a box spits out a ticket containing the terms and conditions for parking there. By passing through the barrier, you acquiesce to such terms and park your vehicle accordingly. What the vehicle owner does not agree to is for the garage owner to operate the vehicle or let someone else operate it as a power source for the garage's electricity without permission or compensation. This is analogous to the claims that Microsoft and GitHub make. *See* Defendants' Motion to Dismiss at 6–7, Doe v. GitHub, Inc., Nos. 4:22-cv-6823, 4:22-cv-7074 (N.D. Cal. Jan. 26, 2022) [hereinafter Defs. Mot.] (rejecting Plaintiff's breach of contract claims). However, simply because the owner parked the vehicle and accepted the terms and conditions of the garage, which did not include use of the engine, the garage owner cannot operate the vehicle without the express consent of its owner. GitHub argues that the plaintiffs failed to specify exactly what provisions of the license agreements they have violated, arguing that "because none of the open-source licenses attached to the Complaint appear to prohibit" training Copilot from public repositories, "the principles embodied in customary open source licenses contemplate broad public rights to inspect, learn from, and build upon code." *Id.* at 13.

[14] *See* Ayala Goldstein, *Open Source Licenses in 2020: Trends and Predictions,* WHITESOURCE (Jan. 23, 2020), https://web.archive.org/web/20200503111426/https://resources.whitesourcesoftware.com/blog-whitesource/top-open-source-licenses-trends-and-predictions (last visited Oct. 1, 2023) (explaining 2020 open-source license trends); *see, e.g.*, Jerome H. Saltzer, *The Origin of the "MIT License,"* IEE ANNALS HIST. COMPUTING 94, 94 (2020) (recounting formation and distribution of MIT License).

[15] *See* Goldstein, *supra* note 14 (commenting that "permissive licenses are winning" at cooperation and minimizing challenges).

[16] *See* Compl., *supra* note 4, ¶ 3 (describing GitHub's formation).

[17] *See id.* ¶¶ 3–4 (alleging GitHub failed to fulfill its promises); *see also* Christopher Tozzi, *What Is GitHub and What Is It Used For?*, ITPRO TODAY (Sept. 9, 2022),

collaboration, GitHub's website recommends that programmers add a "README" file to each of their repositories that inform other users about why it was created or how to use code, while also providing any relevant licensing and attribution guidelines.[18] While most public repositories have an open-source code license stored in each file, those without one are subject to ordinary copyright law.[19]

GitHub and other repositories are, by analogy, the equivalent of a "Digital Library of Congress." It is an infinite landscape of electronic files where each developer's files and revision history are stored, multiplied by the 25 million developers that have registered, uploaded, and parked their open-source code with GitHub since 2008.[20] Either owned individually or shared by an organization, a repository is a single software project stored on free internet servers which allow developers from anywhere around the world to collaborate on public open-source code.[21] Repositories can also be programmed to be private and hidden from the public eye, which are primarily used by organizations seeking to restrict and limit access

---

https://www.itprotoday.com/devops/what-github-and-what-it-used [https://perma.cc/7XWL-QDNN] (describing GitHub's core features).

[18] *See About READMEs*, GITHUB, https://docs.github.com/en/repositories/creating-and-managing-repositories/best-practices-for-repositories [https://perma.cc/XW9V-GHNU] (last visited Sept. 1, 2023) (providing user guidance for setting up repositories).

[19] *See* Compl., *supra* note 4, ¶ 119 (explaining alternatives to open-source licenses).

[20] *See About Repositories*, GITHUB, docs.github.com/en/repositories/creating-and-managing-repositories/about-repositories [https://perma.cc/785E-KJZ6] (last visited Sept. 1, 2023) (defining the scope of repositories); *see also* Compl., *supra* note 4, ¶ 3 (providing GitHub user statistics); *see also* Tozzi, *supra* note 17 (reporting that as of 2022, GitHub had attracted 83 million users).

[21] *See* Compl., *supra* note 4, ¶¶ 112–15 (describing the intent behind open-source platforms).

through password protection.[22] At their core, however, repositories are designed to be used by the public.[23]

Microsoft acquired GitHub in October 2018 for $7.5 billion on the mantra, "Microsoft Loves Open Source."[24] However, by investing $1 billion in OpenAI LP (a for-profit subsidiary of the nonprofit OpenAI), in 2020, Microsoft became the exclusive licensee of OpenAI's GPT-3 language model.[25] OpenAI's trajectory was fixed in 2015 when a group of researchers established a "non-profit artificial intelligence research company" that shared members of its board of directors with Microsoft and was chaired by the likes of Elon Musk.[26] In 2016, Microsoft partnered with OpenAI to build a supercomputer in Microsoft's Azure cloud-computing platform to train its AI models on code completion, which eventually became the underlying model for its AI coding assistant "Copilot"—earning Microsoft the title, "unofficial owner of OpenAI."[27]

GitHub (now owned by Microsoft) was included in *J. Doe v. GitHub, Inc.* because it allegedly stole source code from its members without any owner attribution or licensing agreements in order to develop Copilot.[28] Named defendant OpenAI was also included in this action for its product "Codex," a system that converts natural language into code that is integrated into Copilot, which enables users to prompt Copilot with human language to receive a Codex-suggested solution.[29] Copilot was unveiled to the public in June 2022 as an assistive AI-based system that emits a "possible completion of . . . code" ("output") when software programmers prompt it with incomplete code snippets.[30] The plaintiffs in *J. Doe v. GitHub, Inc.* also allege that OpenAI is partially owned by Microsoft (officiating Microsoft's "unofficial" ownership status, as mentioned above).[31]

---

[22] *See About Repositories*, *supra* note 20 (distinguishing repository visibility); *see also* Kyle Wiggers, *Twitter Reveals Some of Its Source Code, Including Its Recommendation Algorithm*, TECHCRUNCH (Mar. 31, 2023), https://techcrunch.com/2023/03/31/twitter-reveals-some-of-its-source-code-including-its-recommendation-algorithm/ [https://perma.cc/9LDW-KCUC]. Last week Twitter published two repositories on GitHub containing their source code relating to how Twitter controls what tweets are generated on a user's "For You" timeline. *Id.* In a move towards being "more transparent," Twitter also hoped revealing their repositories would allow for mistakes in the code to be caught and corrected. *Id.*

[23] *See* Gralla, *supra* note 3 (noting open-source code isn't in the public domain but becomes available for use by the public via licensing terms).

[24] *See* Compl., *supra* note 4, ¶ 121 (detailing GitHub's growth).

[25] *See id.* ¶ 128 (alleging Microsoft became part owner of GitHub). GPT-3 is a language model that allows AI technology to produce "naturalistic text." *See id.* ¶ 131. When researchers realized GPT-3 could also generate software code, OpenAI and Microsoft began to develop Codex, a code-completion model that became the basis for Copilot. *See id.* ¶¶ 131–32.

[26] *See id.* ¶¶ 122, 124 (detailing the formation of OpenAI).

[27] *See id.* ¶¶ 130, 132 (describing the Microsoft/OpenAI collaboration).

[28] *See* Jasmin Jackson, *Microsoft, Others Want Out of Source Code Copyright Suit*, LAW360 (Jan. 27, 2023), https://www.law360.com/articles/1570223/microsoft-others-

Launched as a subscription-based service for a fee of either $10 per month or $100 per year, Copilot's output is derived from and trained by existing code in public repositories on GitHub.[32] GitHub and Microsoft have described the scope of its Copilot program as "a coding assistant tool that crystallizes the knowledge gained from billions of lines of public code, harnessing the collective power of open source software and putting it at every developer's fingertips."[33] The code training enables the software to detect statistical patterns rather than engage in human-like reasoning.[34] The plaintiffs in *J. Doe v. GitHub, Inc.* allege that Copilot is ingesting these billions of lines of code without being trained to identify the owner of the code, all

---

want-out-of-source-code-copyright-suit [https://perma.cc/8SLS-XDX9] (recounting the motions to dismiss filed by the defendants).

[29] *See* Compl., *supra* note 4, ¶¶ 130–31 (recounting the creation of the generative AI model); *see also* Thomas Maxwell, *Developers Are Turning To GitHub Copilot, a ChatGPT-like Tool That Helps Them Write Code. One Startup VP Says It Helped Him Save 10% of the Time He'd Spend Coding,* BUS. INSIDER (Mar. 6, 2023), https://www.businessinsider.com/codex-github-copilot-chatgpt-openai-productivity-2023-3 [https://perma.cc/JB6X-339D] (commenting on the excitement surrounding Copilot's accessibility).

[30] *See* Compl., *supra* note 4, ¶¶ 45–46 (describing the mechanism by which Copilot generated output).

[31] *See id.* ¶¶ 128, 130–31 (speculating that by investing $1 billion into OpenAI, Microsoft acquired part ownership).

[32] *See id.* ¶¶ 8, 22 (alleging Copilot used public source-code without permission).

[33] *See* Defs.' Mot., *supra* note 13, at 1 (announcing the purpose behind developing Copilot).

[34] *See* Compl., *supra* note 4, ¶ 81 (distinguishing AI reasoning from human-like reasoning).

while failing to provide attribution, copyright notices, or license terms attached to the output.[35]

The plaintiffs in *J. Doe v. GitHub, Inc.* assert numerous causes of action, including breach of contract under GitHub's Privacy Statement, GitHub's Terms of Service, and other various licenses.[36] The plaintiffs further allege that by accepting GitHub's Terms of Service, they formed a contract that included a promise by GitHub not to sell or distribute licensed materials outside of GitHub.[37] GitHub allegedly breached these representations by operating Copilot, which "stripped" the source code of its attribution and license terms and distributed the "now-anonymized code to Copilot users as if it were created by Copilot."[38]

## B. The Need for the "Human Touch"

A review of the pleadings, even at the early stage of this litigation at the time of this Article's writing, shows how the emergence of AI systems and its correlation with copyright law raises a multitude of issues, particularly concerning the dynamic between human authorship and AI-generated work.[39] There are three potential ways for U.S. copyright law to navigate the difficulties in attributing works created by AI where there is little to no human involvement:[40]

---

[35]    *See id.* ¶¶ 56, 82–83 (listing causes of action). The anonymous plaintiffs assert that Copilot was not programmed to "treat attribution, copyright notices, and license terms as legally essential" and that the defendants made a "deliberate choice" to accelerate its 2021 launch instead of prioritizing legal compliance. *Id.* ¶ 80.
[36]    *See id.* ¶ 1 (stating basis of complaint). GitHub's Terms of Service and Privacy Statements promise to not sell or distribute Licensed Materials outside GitHub, but it is alleged to have shared Licensed Materials on Copilot, an external extension that is not part of GitHub. *See id.* ¶¶ 191–93.
[37]    *See id.* ¶¶ 212–13, 216 (asserting GitHub's misrepresentations breached open-source license agreements).
[38]    *See id.* ¶ 11 (claiming GitHub's unauthorized adoption of plaintiffs' source code resulted in disguising its developers).
[39]    *See* Jillian M. Taylor, *AI and Copyright: A New Kind of Copyright Troll? The Rise of AI in Creative Works*, ALM (Mar. 22, 2023, 10:55 AM), https://www.law.com/thelegalintelligencer/2023/03/22/ai-and-copyright-a-new-kind-of-copyright-troll-the-rise-of-ai-in-creative-works/ [https://perma.cc/KAD5-VFZR] (describing AI copyright implications).
[40]    *See Copyright Office Launches New Artificial Intelligence Initiative*, U.S. COPYRIGHT OFF. (Mar. 16, 2023), copyright.gov/newsnet/2023/1004.html [https://perma.cc/3LST-

(1) regulators can outright deny protection for work generated by AI programs; (2) regulators can grant copyright protection and attribute authorship to the programmer who created the AI program; or (3) grant copyright protection and attribute authorship to the end user who provided information to the AI program to generate the resulting work.[41] However, under any of these circumstances, there is likely a human developer excluded from copyright protection that generated the source code used to either create the resulting work or form the basis of a generative AI program's output.

Moreover, the U.S. Copyright Office ("USCO") recently drew a line in the sand stating that there must be human involvement in the authorship of an AI-generated work to acquire copyright protection.[42] Courts interpreting the phrase "works of authorship" have uniformly limited it to the creations of human authors.[43] In cases where non-human authorship is claimed, appellate courts have found that copyright does not protect the alleged creations.[44] Using the same analysis, the USCO refused to register a two-dimensional artwork titled "A Recent Entrance to Paradise," which was created autonomously by the AI system DABUS, because it "lack[ed] the human authorship necessary to support a copyright claim."[45] The case is currently before the D.C. District Court, where Stephen

---

ZWKQ] (announcing that on March 16, 2023, the USCO launched "[a] new initiative to examine the copyright law and policy issues raised by artificial intelligence (AI), including the scope of copyright in works generated using AI tools and the use of copyrighted materials in AI training.").

[41] *See* Taylor, *supra* note 39 (proposing three remedies to tackle copyrightability of AI-generated work).

[42] *See id.* (referencing the "Zarya of the Dawn" decision).

[43] Letter from Robert J. Kasunic, Assoc. Reg. of Copyrights and Dir. of Off. Registration Pol'y and Prac., U.S. Copyright Off., to Van Lindberg, Taylor English Duma LLP (Feb. 21, 2023) [hereinafter "Zarya of the Dawn Correspondence"], https://www.copyright.gov/docs/zarya-of-the-dawn.pdf [https://perma.cc/65Q5-WKQF].

[44] *Id.* at 4.

[45] *See* Letter from Shira Perlmutter, Reg. of Copyrights, U.S. Copyright Off. Rev. Bd., to Ryan Abbott, Brown, Neri, Smith & Khan, LLP (Feb. 14, 2022), https://www.copyright.gov/rulings-filings/review-board/docs/a-recent-entrance-to-paradise.pdf [https://perma.cc/AB3T-J93Q] (affirming registration denial for "A Recent Entrance to Paradise"). Stephen Thaler created DABUS. *See* ARTIFICIAL INVENTOR PROJECT, https://artificialinventor.com/ [https://perma.cc/8E5K-AF2U] (last visited Apr. 16, 2023).

Thaler, the plaintiff and DABUS' developer, argued: "The plain language of the Copyright Act ('Act') clearly allows non-human authors. Nor does anything in the Act exclude certain non-human authors, such as AI systems, from creating copyrighted works."[46] The USCO rejected this argument:

> AI is an emerging technology, and neither Plaintiff nor Defendants are aware of cases specifically addressing whether AI can be considered an author under the Act. Appellate Courts have, however, considered analogous cases regarding works created by animals, nature, and other non-humans. These decisions employ reasoning that applies squarely to this case and have uniformly rejected non-human authorship of copyrighted works.[47]

## C. The "Fair Use" Defense

What may be the downfall for programmers or developers seeking copyright protection for AI-generated works, such as artwork or visual images, is the affirmative defense of fair use. The fair use doctrine provides that unauthorized use may be permissible if it builds upon works in a manner that does not deprive copyright owners of the right to control and benefit from such works.[48] GitHub and Microsoft used the same arguments in their motion to dismiss in *J. Doe v. GitHub, Inc.*:

> Plaintiffs fail to identify any of the "Licensed Materials" they allegedly placed in a GitHub public repository that reflect purported "copyright interests," or to tell us anything at all about those materials. The Complaint nowhere identifies any copyrighted work owned by either of the Plaintiffs, or any registration of such work. The Complaint fails to identify any use of their Licensed Materials. Although the case is supposedly about "software piracy on an unprecedented scale," Plaintiffs make no copyright infringement claim. And, Plaintiffs identify no personal identifying information that they stored in their public repositories on GitHub, or say how it was allegedly exposed by Codex or Copilot.[49]

---

[46] Plaintiff's Combined Opposition to Defendant's Motion for Summary Judgement & Reply in Support of Plaintiff's Motion for Summary Judgement at 1, Thaler v. Perlmutter, 2023 WL 5333236, at *1 (D.D.C. Mar. 7, 2023) (No. 22-1564) (criticizing the U.S. Copyright Office's (USCO) imposition of a human authorship requirement).

[47] Defendants' Reply in Support of Their Cross-Motion for Summary Judgement at 8–9, *Thaler*, 2023 WL 5333236, at *1 (D.D.C. Apr. 5, 2023) (citations omitted) (affirming its reasoning that non-human authorship is not qualified under copyright law).

[48] *See* Taylor, *supra* note 39 (explaining the fair use affirmative defense).

According to GitHub and Microsoft, the anonymous plaintiffs in *J. Doe v. GitHub, Inc.* did not claim copyright infringement because they wanted "to evade the limitations on the scope of software copyright and the progress-protective doctrine of fair use."[50] In the same context, GitHub and Microsoft also claimed that training AI systems on publicly available code is a widely accepted practice of "fair use."[51] To add another wrinkle to this novel area, no court has considered the question of whether "training machine learning models on publicly available data is . . . fair use."[52]

## D. Humans vs. Machines

Current AI systems trained in code are nothing more than processors that can suggest or simulate statistical patterns, which is certainly not the equivalent of human-like reasoning.[53] Whether AI systems will make humans lazy, hinder our ability to think, or eradicate the need for creativity are issues that will be debated for years. What is clear now is that AI systems will never be lazy, as they will operate continuously at the push of a button. Creativity, on the other hand, is not a statistical configuration of digits and probabilities. It is this divergent approach to creativity that may conflict with the Copyright Act, which limits the scope of copyright-eligible matter to "original works of authorship fixed in any tangible medium of expression."[54] The United States Supreme Court has held that the term "original" consists of two components: independent creation and sufficient creativity.[55] That is, the protected work must have been independently created by the author.[56] Second, the work must possess sufficient creativity.[57] In the context of AI-generated works, many questions arise. Is "sufficient creativity" satisfied by AI systems that suggest or simulate statistical patterns? Are AI systems "capable of 'simulat[ing] human reasoning or inference,' or can they engage in the same sort of pattern recognition, synthesis, and prediction" as humans?[58] Do such patterns rise to the standard under the Copyright Act of possessing "the inventive or master mind"?[59]

---

[49] *See* Defs. Mot., *supra* note 13, at 5 (asserting a fair use defense).
[50] *See id.* at 1 (claiming the plaintiffs strategically avoided copyright infringement claim to circumvent the fair use defense).
[51] *See* Compl., *supra* note 4, ¶ 84 (challenging defendants' claims). GitHub has asserted that "training for Codex (the model used by Copilot) is done by OpenAI, not GitHub" and justified its use of copyrighted code as training data on the premise that "computational analysis and training of machine learning models . . . do not require consent of the owner of such materials. Such laws are intended to . . . ensure public benefit" of the material. *Id.*
[52] *See id.* (recognizing the unique procedural posture). The District Court ruled on the Motion to Dismiss on May 11, 2023, dismissing most of the Plaintiffs' claims with leave to amend. The District Court denied the motion to dismiss the alleged violations of Sections 1202(b)(1) and 1202(b)(3) of the Digital Millennium Copyright Act (DMCA). *See* Mana Ghaemmagham & Stuart Levi, *Ruling on Motion To Dismiss Sheds Light on Intellectual Property Issues in Artificial Intelligence*, SKADDEN, ARPS, SLATE, MEAGHER & FLOM LLP (May 24, 2023), https://www.jdsupra.com/legalnews/ruling-on-motion-to-dismiss-sheds-light-6984451/ [https://perma.cc/W8ND-EPRP].

All in all, the prerequisites of human authorship and "sufficient creativity" within the context of the U.S. Copyright Act will not be easily reconciled with the defense of fair use if AI systems simply suggest or simulate statistical patterns—and even that remains yet to be determined.

---

[53]　*See* Compl., *supra* note 4, ¶ 81 (explaining AI functionality, that "AI models cannot 'learn' as humans do, nor can it 'understand' semantics and context the way humans do. Rather, it detects statistically significant patterns in its training data and provides Output derived from its training data when statistically appropriate"); *see also DABUS Described*, IMAGINATION ENGINES, https://imagination-engines.com/dabus.html [https://perma.cc/2LBX-ST2Z] (last visited Apr. 16, 2023). However, Stephen Thaler claims that his machine "DABUS" can create a stream of consciousness. *Id.*

[54]　*See* 17 U.S.C. § 102(a).

[55]　*See* Feist Publ'ns, Inc. v. Rural Tel. Serv. Co., 499 U.S. 340, 345 (1991).

[56]　*See id.*

[57]　*See id.*

[58]　*See* Defs. Mot., *supra* note 13, at 1–2.

[59]　*See* Burrow-Giles Lithographic Co. v. Sarony, 111 U.S. 53, 61 (1884).

I. DISCUSSION

A. *The Contractual Enforcement Battle: Licensing Agreements vs. Terms of Service*[60]

Licensors of open-source code stored in public repositories are in a unique position in that they have an exclusive copyright by default, but often expect that their code will be used and advanced by others.[61] Therefore, they require an open-source code license to allow others to use, modify, and share their work.[62] As mentioned above, some of the most popular open-source licenses include the MIT License (*see Exhibit 1*) and GPLv3, both of which provide that the open-source code may only be used with express permission from the licensor.[63] Further, GitHub explicitly notes that creating a public repository of open-source code without including a license does not automatically give others the right to use it without obtaining the necessary permissions:

> Making your GitHub project public is not the same as licensing your project. Public projects are covered by GitHub's Terms of Service, which allows others to view and fork your project, but your work otherwise comes with no permissions.
> If you want others to use, distribute, modify, or contribute back to your project, you need to include an open source license. For example, someone cannot legally use any part of your GitHub project in their code, even if it's public, unless you explicitly give them the right to do so.[64]

---

[60] We believe the license terms of the open-source code should override any conflicting Terms of Service by repositories. We think of this as the terms of access to a Library, which should not override the copyright of authors that have their books stored in the Library, which the Library then shares with its members. *See supra* note 18 and accompanying text.
[61] *See The Legal Side of Open Source*, OPEN SOURCE GUIDES, https://opensource.guide/legal/ [https://perma.cc/X3HD-ZYPS] (last visited Aug. 27, 2023) (explaining the "legal implications of open source").
[62] *See id.*
[63] *See id.*
[64] *See id.*

A problematic dynamic arises when the open-source code is licensed but a software program's Terms of Service or use contradict the applicable license. This situation is exemplified in Microsoft's Visual Studio Code Software License Terms, which could displace or contravene the terms of an open-source code license agreement.[65] Specifically, the Terms of Service for Microsoft's code editor software states that open-source code is licensed under the MIT license agreement, which provides that permission is granted "to any *person* obtaining a copy of this software . . . to use, copy, modify, merge, publish, distribute, sublicense, and/or sell copies of the Software . . . ."[66] However, Microsoft's Terms of Service do not provide the same guarantee that access to the software is limited to *human* access, but instead broadly states that "you may use the software only as expressly permitted in this agreement."[67] This allows for a scenario wherein Generative AI co-opts licensed open-source code, leaving developers that code forced to consider litigation or simply suffer the intellectual property loss.[68]

Furthermore, the open-source code's presence in a public repository on GitHub, coupled with the code's owner having signed a separate terms and conditions acknowledgment to use GitHub's platform, could lend itself to the code being trained into Generative-AI models, thus potentially breaching a separate license that restricts use to "person[s]" only.[69] In *J. Doe v. GitHub Inc.*, the plaintiffs allege that GitHub violated its Terms of Service by training Copilot on open-source code, accusing GitHub of having "held itself out as the best place to host open-source code repositories," but failing to honor such representations.[70] This poses the dilemma of which agreement governs and whether more clearly defined limitations prohibiting non-human access in license agreements could overcome this conflict.[71] Further, the issue of whether the open-source code license or the platform's Terms of Service agreements governs is exacerbated by ambiguity as to whether these agreements are preempted by the Copyright Act altogether.[72]

---

[65] *See Microsoft Software License Terms*, VISUAL STUDIO CODE, https://code.visualstudio.com/license [https://perma.cc/X5CG-GB9Q] (last visited Aug. 27, 2023) (listing applicable licensing terms).

[66] *See microsoft/vscode/LICENSE.txt*, GITHUB (emphasis added), https://github.com/microsoft/vscode/blob/main/LICENSE.txt [https://perma.cc/2CHS-V8ZT] (last visited Aug. 27, 2023) (showing an attributable software license).

[67] *See Microsoft Software License Terms*, *supra* note 65 (contradicting attributable license terms); *see also Microsoft Visual Studio Marketplace Terms of Use*, MICROSOFT, https://cdn.vsassets.io/v/M190_20210811.1/_content/Microsoft-Visual-Studio-Marketplace-Terms-of-Use.pdf [https://perma.cc/97LD-8T2R] (last updated June 2021) (providing Microsoft's Terms of Use).

[68] *See* Compl., *supra* note 4, ¶ 15. This constitutes a hypothetical intellectual property and profit loss because generative AI "monetize[s]" other programmers' code instead of providing attribution or even compensation in certain circumstances. *Id.*

[69] *See id.* ¶ 212 (asserting that Plaintiffs accepted GitHub's Terms of Service promising not to sell licensed materials); *see also Microsoft Software License Terms*, *supra* note 65 (lacking the language limiting use to human persons only).

Circuit courts are split on whether breach-of-contract claims arising from a contractual promise regarding copyrighted material are preempted by the Copyright Act, such as when a licensee violates an open-source code license agreement.[73] The nature of the split involves whether (1) a contractual promise not to copy copyrighted material will be treated as a breach of contract under state law or (2) it will be governed by the remedies available under the Copyright Act.[74] The Fifth, Seventh, and Eleventh Circuits have been proponents of the view that contractual promises avoid preemption, reasoning that a copyright puts the world on notice, whereas a contractual agreement is specific to the parties entering such an agreement.[75] In contrast, the Second and Sixth Circuits examine the issue of whether the contractual rights are "qualitatively different" from the rights secured by the Copyright Act on a case-by-case basis.[76] The Eighth Circuit has indicated that it is in line with the Fifth, Seventh, and Eleventh circuits' approach by holding that a contractual agreement restricting the use of a licensed program constitutes an "extra element" sufficient to distinguish contractual rights from a copyright action.[77] This continuing circuit split could be subject to Supreme Court review and serves as a warning to contractual parties to include forum selection clauses in licenses if they wish to achieve a particular outcome.[78]

---

[70] *See* Compl., *supra* note 4, ¶¶ 192–93 (criticizing GitHub's public image).
[71] This is our recommendation. *See infra* Exhibit 2.
[72] *See* Arlene Boruchowitz & Jaci Overmann, *With End-User License Agreements, Which Will Prevail: Copyright Rights or Contract Rights?*, DINSMORE (Nov. 28, 2022), https://www.dinsmore.com/publications/with-end-user-license-agreements-which-will-prevail-copyright-rights-or-contract-rights/#_edn8 [**https://perma.cc/HS62-E74D**] (explaining the nature of the federal circuit split over copyright preemption).
[73] *See id.* (clarifying the circuit split).
[74] *See id.* ("The Fifth, Seventh, and Eleventh Circuits have all suggested that a contractual promise itself is an 'extra element' sufficient to avoid preemption," and the Sixth and Second Circuit apply a more flexible rule that examines on a case-by-case basis whether the specific contractual rights are significantly different from the Copyright Act's exclusive rights.").
[75] *See id.* (reporting a favorable contractual interpretation).

A promising sign for developers of open-source code is the recent decision in *Software Freedom Conservancy, Inc. v. Vizio, Inc.*, where the U.S. District Court for the Central District of California allowed a consumer to proceed on a breach of contract claim where a product maker failed to share open-source code software in contravention of its open-source license agreements.[79] There, Software Freedom Conservancy, Inc. ("SFC") purchased smart TVs from Vizio Inc. ("Vizio") and brought a suit alleging breach of contract after Vizio failed to provide source code to compile the devices' software.[80] SFC alleged it was adversely affected when Vizio violated two general public license ("GPL") agreements which required that "those who distribute software in an executable form . . . also make the software available as 'source code,' . . . thus allowing [consumers] to further develop the software."[81] The *Vizio* court held that the claim was not preempted by copyright law.[82] While this was not specific to the repositories exemplified in GitHub, the court's holding is significant for potentially extending breach of open-source code licenses under a contractual theory.[83]

---

[76]   *See id.* (delineating a more rigid interpretation).

[77]   *See id.*; *see also* Nat'l Car Rental Sys., Inc. v. Comput. Assocs., 991 F.2d 426, 431 (8th Cir. 1993).

[78]   *See* Boruchowitz & Overmann, *supra* note 72 (noting inconsistent case law); *see infra* Exhibit 2, n. 155 (proposing a forum selection clause).

[79]   *See* Software Freedom Conservancy, Inc. v. Vizio, Inc., No. 8:21-cv-01943, 2022 U.S. Dist. LEXIS 87115, at *13 (C.D. Cal. May 13, 2022); *see also Microsoft Software License Terms*, *supra* note 67; Jeremy Elman, *Vizio Ruling Offers Potential Precedent On Source Code*, LAW360 (July 8, 2022, 3:40 PM), https://www.law360.com/articles/1506538/vizio-ruling-offers-potential-precedent-on-source-code [https://perma.cc/HZM9-2RXQ] (summarizing holding).

[80]   *See* Elman, *supra* note 79.

[81]   *See Vizio,* 2022 U.S. Dist. LEXIS 87115, at *2–3 (citing plaintiff's reasoning); *see also* Katie Terrell Hanna, *GNU General Public License*, TECHTARGET, https://www.techtarget.com/searchdatacenter/definition/GNU-General-Public-License-GNU-GPL-or-simply-GPL [https://perma.cc/8JRL-CQHX] (last visited Sept. 19, 2023) (recounting license developments to combat proprietary software).

Another development is a petition for a writ of certiorari filed in *ML Genius Holdings LLC v. Google LLC*.[84] There, ML Genius Holdings ("Genius"), a website that licenses the right to transcribe and display lyrics from copyrighted music, sued Google for allegedly violating Genius' terms and conditions by copying Genius' content for commercial purposes.[85] Genius alleged that Google accepted the website's terms but then stole the lyrics for use on its own competing site.[86] The Second Circuit held that Genius' claim that it protected its content through its Terms of Service was preempted by the Copyright Act because Genius had not demonstrated how its claims were "qualitatively different" from a copyright claim for lyrics it did not own.[87] Commentators have noted that "Genius's case appears to be the perfect vehicle for the Supreme Court" to decide which test will govern these issues.[88] This body of case law suggests that private agreements may or may not be preempted by the Copyright Act depending on the circuit where the suit is brought (pending the Supreme Court granting certiorari) but is a positive sign for licensors of open-source code agreements hoping to bring contractual claims against Generative AI programs.[89]

---

[82] *See Vizio,* 2022 U.S. Dist. LEXIS 87115, at *8–9 (holding that the extra element entitled SFC to receive source code under their agreements).

[83] *See* Elman, *supra* note 79 (reporting Vizio's contractual implications).

[84] *See* ML Genius Holdings LLC v. Google LLC, No. 20-3113, 2022 U.S. App. LEXIS 6206 (2d Cir. Mar. 10, 2022).

[85] *See id.* at *1; *see also* Petition for Writ of Certiorari, ML Genius Holdings LLC v. Google LLC, 143 S.Ct. 2658 (No. 22-121); Boruchowitz & Overmann, *supra* note 72, at 4 (discussing the likely consequence of the Supreme Court's decision to deny certiorari).

[86] *See* Boruchowitz & Overmann, *supra* note 72; *see also* Tiffany Hu, *Copyright & Trademark Cases to Watch in 2023*, LAW360 (Jan. 2, 2023), https://www.law360.com/articles/1558959/copyright-trademark-cases-to-watch-in-2023 [**https://perma.cc/SN88-4CNY**] (explaining the potential impact if reviewed by Supreme Court).

[87] *See MLGenius Holdings LLC*, 2022 U.S. App. LEXIS 6206, at *1, 11 (holding that Genius failed to plead the extra element to differentiate it from a copyright claim).

[88] Boruchowitz & Overmann, *supra* note 72 (calling for the Supreme Court to weigh in on the circuit split).

## B. A Call to Action

This dichotomy in judicial interpretations underscores the question of what should come first: innovation or legislation. "Legislating emerging technologies is challenging when there is not a full understanding of all the implications and nuances and there is no social activity other than what is being done in the lab. Such issues may not be ripe for legislative resolution or social trendsetting."[90] Despite this social reluctance, we believe that the lack of legislative framework for emerging technologies hinders and delays innovation, particularly in AI systems. Even further, it is critical that the technological direction of AI systems (an important innovation milestone) not be delegated to the hands of a few super-powerful conglomerates because it would directly contradict the humanity-first nature of open-source code:

It's an extraordinary idea, one that has fostered an immense body of public knowledge, ever-evolving and ever-available for the next generation of developers to build, collaborate, and progress. The open source model is essential to collaborative and communal software development. GitHub was founded on the basis of these ideals, and when Microsoft invested billions to acquire GitHub in 2018, it cemented its commitment to them. The transformative technology at issue in this case, Copilot, reflects GitHub and Microsoft's ongoing dedication and commitment to this profound human project.[91]

---

[89] *See id*; *see also* Thomas Claburn, *Voice.ai Denies Claim it Violated Open Source Software License Requirements*, REGISTER (Feb. 8, 2023), https://www.theregister.com/2023/02/08/voiceai_open_source/ [https://perma.cc/8AVC-K9YU] (reporting potential violations of open-source code license agreements). An interesting development whereby a software developer and security researcher known as Ronsor wrote a blog post after discovering that Voice.ai, maker of a voice-changing application, violated two open-source licenses in its libraries. *Id.* Ronsor reports Voice.ai used source-code from two third parties in its voice-changing software but failed to provide any attribution by including the licenses with the software. *Id.* Voice.ai's Terms of Service, which expressly forbids the "copying, modification, and reuse of the software," also violate the licenses and directly contradicts the open-source licenses "that require those freedoms." *Id.*; *see also* Dean Howell, *Jailbreak Hacker Uncovers 'Stolen' Open-Source CodeHacker in Voice.ai*, NEOWIN (Feb. 6, 2023), https://www.neowin.net/news/jailbreak-hacker-uncovers-stolen-open-source-code-in-voiceai/ [https://perma.cc/MBS7-J8TC] (recounting Ronsor's interaction with Voice.ai). "Misuse of open-source software can threaten the integrity of the open-source community and undermines the principles that make open-source software so valuable." *Id.* Interestingly, Ronsor communicated his discovery to Voice.ai and was then banned from Voice.ai's Discord server shortly thereafter. *Id.*

[90] Dimitrios Ioannidis, *Will Artificial Intelligence Replace Arbitrators Under the Federal Arbitration Act?*, 28 RICH. J.L. & TECH. 505, 578–79 (2022) (commenting on the frustration in the scientific community inhibiting innovation).

For these reasons, advocating for legislative action would both promote innovation and implement safeguards to eliminate the raiding of open-source code, especially with respect to militaristic AI applications.[92] The implementation of international standards is one of the goals of the DILEMA Project ("Project"), which is an ongoing research project aimed at assessing the ethical and legal implications of AI technology in the military.[93] One particular aim of the Project is to evaluate human involvement in the deployment of such technologies: "It is therefore essential to critically address the questions of where and how the role of human agents should be maintained throughout the development and deployment of military AI technologies."[94] It is this type of imperative work that will maximize AI's functionality while still regulating its ethical and legal ramifications,[95] thus drawing attention to the need for a legislative framework to define its parameters and bolster innovation.

C. *"Is any of this actually legal?" The Fair Use Defense*[96]

In response to the aforementioned suit, GitHub, OpenAI, and others have raised the affirmative defense of "fair use" against the claims that they have violated open-source code licenses.[97] The fair use defense is enshrined in U.S. law and permits the limited unauthorized use of copyrighted material based on four factors: (1) the purpose and character of the use, including whether the use is of a commercial nature or is for nonprofit educational purposes; (2) the nature of the copyrighted work; (3) the amount and substantiality of the portion used in relation to the copyrighted work as a whole; and (4) the effect of the use upon the potential market for or value of the copyrighted work.[98] In the world of AI, fair use is a frequently asserted defense, but its viability in this groundbreaking area remains unknown.[99]

---

[91] Defs. Mot., *supra* note 13, at 1 (challenging the plaintiffs' contentions) (emphasis added).
[92] *See* Taylor Woodcock, *Artificial Intelligence: Is the Answer More Law?*, ASSER INST. BLOG (Jan. 25, 2021), https://www.asser.nl/about-the-asser-institute/news/blog-artificial-intelligence-is-the-answer-more-law/ [https://perma.cc/TE5D-K72E].
[93] *See Research Project*, DILEMA, https://www.asser.nl/dilema/about-the-project/research-project/ [https://perma.cc/24QM-6HLV] (last visited Sept. 2, 2023).
[94] *Id.*
[95] *See id.* (questioning the broader application of AI technology).
[96] *See* James Vincent, *The Scary Truth About AI Copyright Is Nobody Knows What Will Happen Next*, VERGE (Nov. 15, 2022), https://www.theverge.com/23444685/generative-

The fair use defense is assessed on a case-by-case basis, thus posing challenges for a licensor (i.e., copyright holder) looking to defeat the defense.[100] The defense is premised on a policy of promoting freedom of expression, but in deciding the purpose and character of the use, courts look to whether the use changes the nature of the material in some way, commonly referred to as a "transformative use."[101] The line between what is transformative and what is not could potentially hinge on whether Generative AI is using the copyrighted data to generate its output in a manner that changes the material (likely constituting sufficient transformation) or whether it is using the data to train its system to model the data itself (likely not constituting transformation).[102] We believe in implementing a policy that trains generative AI systems to recognize and incorporate the rights of open-source developers because it will lead to safer and more accurate systems for public use.[103] At the same time, we

---

ai-copyright-infringement-legal-fair-use-training-data [https://perma.cc/EWK4-MK35] (remarking on the increasing concerns over AI systems' replication of human data).

[97] *See* Defs. Mot., *supra* note 13, at 88.
[98] 17 U.S.C. § 107; *see* Taylor, *supra* note 39.
[99] *See* Taylor, *supra* note 39 ("Whether the AI companies are infringing on the artists' copyrights . . . to not only train the AI programs but also to generate new art . . . is a complicated question that will need to be settled in the courts.").
[100] *See id.* (describing judicial interpretation of the fair use defense).
[101] *See* Vincent, *supra* note 96 (remarking on important considerations by the courts).
[102] *See id.* (indicating that courts may distinguish between training or transformative usage).
[103] *See* Mark A. Lemley & Bryan Casey, *Fair Learning*, 99 TEX. L. REV. 743, 770–71 (2021) (proposing one policy stance in favor of AI systems).

support the improvement of generative AI platforms and systems in the context of innovation, but not at the expense of replicating copyrighted work. Intent aside, creating trustworthy or advanced systems should not come at the cost of infringing on an individual's protected material or creating distrust in sharing innovative material.[104]

Deciding whether particular AI behavior should be classified as output (transformative) or training (non-transformative) is burdensome and will likely hinge on the Supreme Court's discretion.[105] Generative AI companies are aware of this and are mindful of how to manipulate this to their advantage.[106]

> Another variable in judging [the first factor in the] fair use [analysis] is whether or not the training data and model have been created by academic researchers and nonprofits. This generally strengthens fair use defenses and startups know this. So, for example, Stability AI, the company that distributes Stable Diffusion, didn't directly collect the model's training data or train the models behind the software. Instead, it funded and coordinated this work by academics and the Stable Diffusion model is licensed by a German university. This lets Stability AI turn the model into a commercial service (DreamStudio) while keeping legal distance from its creation.[107]

Stability AI ("Stability") and Midjourney are currently in legal crosshairs as both are accused of training their respective AI imaging software programs by downloading billions of copyrighted images without licensors' permissions.[108] The class action complaint contends that Stability commercially benefited and profited from its use of the copyrighted images, and as a result, harmed the plaintiffs through the loss of potential commissions.[109] In their complaint, the plaintiffs allege that Stability's software, Stable Diffusion, creates derivative works via a mathematical software process based on the copyrighted images used in the training process.[110] Stability AI claims that its actions constitute fair use—and it is likely that it will succeed on this argument, given that it has navigated its way into the educational loophole cited above.[111]

---

[104] *See id.* (noting the difficulties in wholly advocating for unexercised AI usage).
[105] *See* Vincent, *supra* note 96 (citing Petition for Writ of Certiorari, Andy Warhol Foundation for the Visual Arts, Inc. v. Goldsmith, 598 U.S. 508 (2023) (No. 21-869)) ("The Supreme Court doesn't do fair use very often, so when they do, they usually do something major. I think they're going to do the same here . . . And to say anything is settled law while waiting for the Supreme Court to change the law is risky.").
[106] *See id.* (commenting how AI companies have strengthened their fair use defenses).
[107] *See id.* (dubbing this practice "AI data laundering").

Stability is not alone. GitHub and OpenAI have offered "shifting justifications" for the source and amount of code used to train their programs and purport to be exempt under the fair use defense.[112] However, GitHub and OpenAI LP may encounter difficulties if they assert an argument similar to the one used by Stability AI. While Open AI may contend it was founded as a nonprofit organization, and therefore its use of the code was not designed for commercial purposes, Open AI LP *was* formed as a for-profit subsidiary of Open AI.[113] It also accepted a $1 billion investment from Microsoft to develop Codex and the resulting Copilot technologies.[114] Further, because the plaintiffs in *J. Doe v. GitHub, Inc.* allege that Microsoft has an ownership interest in Open AI LP, it may further complicate

---

[108] *See* Complaint at 1, Andersen v. Stability AI Ltd., No. 23-00201 (N.D. Cal. filed Jan. 13, 2023) [hereinafter Stability Compl.].
[109] *See id.* at 2.
[110] *See id.* at 1 (alleging Stability appropriated code without attribution). The plaintiffs also emphasize that prior to this software, users looking for an image "in the style of a given artist" had to pay to commission or license an image from the artist, but Stable Diffusion generates the works without compensation to the artist. *Id.* at 1–2.
[111] *See* Taylor, *supra* note 39 (commenting on the unresolved judicial questions). The question of whether Stability AI will be able to assert the fair use defense if it is training its programs *and* generating new art is one that "need[s] to be settled in the courts." *Id.*; *see also* Vincent, *supra* note 96 (documenting the program's educational versus commercial foundation).
[112] *See* Compl., *supra* note 4, ¶ 14.
[113] *See id.* ¶ 127.
[114] *See id.* ¶ 128.

Open AI LP's attempt to distinguish itself from other for-profit enterprises.[115]

Some contend that the best option would be for AI systems to broadly license the entirety of the data used—code, text, images, etc.—thereby eliminating the need to undergo a fair use analysis in the first instance.[116] However, licensing extensive amounts of data risks impeding and eliminating Generative AI systems altogether because "[g]iven the doctrinal uncertainty and the rapid development of [machine learning] technology, it is unclear whether machine copying will continue to be treated as fair use."[117]

### D. Obfuscation: The Needle in the Haystack[118]

"It's official—AI is now the wild west."[119] Despite the public awe at Generative AI models' capabilities, the "hoovering" mechanism by which AI trains (such as broadly inhaling data from publicly accessible repositories of code, text, images, etc.) has prompted widespread backlash and litigation.[120] As noted above, Microsoft, GitHub, and OpenAI are currently the subject of a class action suit because they trained their generative model, Copilot, "on billions of lines of public code, to regurgitate code snippets without providing credit."[121] GitHub has been accused of outputting "verbatim copies of licensed materials" from the open-source code repository to feed its Copilot software without any attribution and/or copyright notice to licensors, however, "[p]laintiffs and the [c]lass have the difficult or impossible task of proving the Licensed Materials belong to them."[122] This difficulty arises from generative AI's obfuscation of open-source code—imposing multiple layers of security to essentially hide, twist, and scramble the code to the point that it is unintelligible, while still retaining access to the original non-obfuscated code.[123] But what are they hiding? Isn't open-source code premised on the idea that it is available for use by the public so long as it is attributed? It begs the question of how far is too far, and what types of liabilities may be imposed where obfuscation exceeds mere security measures.

---

[115] *See id.* ¶ 7.
[116] *See* Vincent, *supra* note 96; *see also* Lemley & Casey, *supra* note 103, at 748.
[117] *See* Vincent, *supra* note 96 (referencing the plausibility of licensing data); *see also* Lemley & Casey, *supra* note 103, at 746–48 (posing difficulties in overcoming the volume of information and surmising that courts will be less sympathetic to AI models being trained with copyrighted material).
[118] *See* Michael Haephrati & Ruth Haephrati, *Who Moved My Code? An Anatomy of Code Obfuscation*, INFOQ (Nov. 9, 2022), https://www.infoq.com/articles/anatomy-code-obfuscation/ [https://perma.cc/HKD3-F8GY] (analogizing that obfuscation is "hiding the needle in the haystack," and that, "if done well, it will take an unreasonable amount of time and resources for an attacker to find your 'needle.'").
[119] *See* Howell, *supra* note 89 (contending that "tech giants" are engaged in "all out warfare" for "AI supremacy").
[120] *See* Gralla, *supra* note 3 (describing AI's adoption of open-source code).
[121] *See* Kyle Wiggers, *The Current Legal Cases Against Generative AI Are Just the Beginning*, TECHCRUNCH (Jan. 27, 2023), https://techcrunch.com/2023/01/27/the-current-legal-cases-against-generative-ai-are-just-the-beginning/ [https://perma.cc/83PU-2UMB].

The difficulty in assessing these questions is that obfuscation is a widely accepted practice in the public online community, which makes determining liabilities all the more challenging. Obfuscation is intended to protect the intellectual property of open-source code and minimize the risk of reverse-engineering.[124] While doubt has been cast on the extent to which substantial obfuscation is reasonable, there is little to suggest that excessive obfuscation has encountered any legal ramifications.[125]

---

[122] *See* Compl., *supra* note 4, ¶ 155.

[123] *See* Haephrati & Haephrati, *supra* note 118 (explaining obfuscation's methodology).

[124] *See* Mark Clement, *Source Code Obfuscation: What it is and Techniques to Use*, STOP SOURCE CODE THEFT (Apr. 10, 2019), https://www.stop-source-code-theft.com/source-code-obfuscation-what-it-is-and-techniques-to-use/ [https://perma.cc/TN58-9ZDC] (explaining obfuscation's protection against risk of reverse engineering); *see also* Renee Zmurchyk, *Contractual Validity of End User License Agreements*, 11 APPEAL: REV. CURRENT L. & L. REFORM 57, 64–66 (2006) (identifying that courts have been inclined to protect against reverse engineering). Reverse engineering is the process of beginning with a finished product and working backwards to figure out how the product was made and how it operates. *Id.* at 64.

[125] *See* Claburn, *supra* note 89 (questioning Voice.ai's obfuscation). Ronsor not only questions the violation of the licenses and Voice.ai's Terms of Service, but also questions the application's "heavy use of obfuscation and the data it collects." *Id.*; *see also Voice.AI: GPL Violations with a Side of DRM*, UNDELETED FILES (Feb. 4, 2023), https://undeleted.ronsor.com/voice.ai-gpl-violations-with-a-side-of-drm/ [https://perma.cc/K3L9-Q797] (criticizing violations of source code licenses by way of obfuscating code). "Voice.ai developers claim that such obfuscation is necessary in order to protect their proprietary secrets (which, by the way, are not allowed to be secrets due to the included GPL code)." *Id.* "[N]o other class of software is this heavily obfuscated,

Questions regarding AI's obfuscation have begun to surface in the legal context. For example, the suit against Microsoft's Copilot and OpenAI's Codex platforms claims that obfuscation disregards the rights associated with licensed materials and suggests that obfuscation is problematic because it fails to attribute licensed material in AI outputs.[126] By generating a derivative work compiled from open-source code licenses, obfuscation allows generative AI to reproduce code from vast repositories and concurrently omit any copyright licensing or attribution to those licenses.[127] Whether the use of obfuscation will be reassessed and liabilities imposed in the course of this litigation remains to be seen.

### E. *Shifting the Burden of Proof is a Possible Remedy to Deal with Obfuscation*

Under the Article III case-or-controversy requirement in the U.S. Constitution, a plaintiff must have standing to bring an action in federal court.[128] The courts have construed this requirement to mean that a plaintiff bears the burden of showing they have suffered an actual or imminent "injury in fact" that is concrete and particularized, and must demonstrate that the injury is traceable to the defendant's actions (i.e., show causation).[129] The causation requirement has posed particular challenges where the type of injury is "untraceable," such as in the case of public shares sold on the New York Stock Exchange.[130] Some courts have sought to address this complication by shifting the burdens of production or persuasion to the defendant. For instance, in *Pirani v. Slack Technologies*, the Ninth Circuit held that a plaintiff need not prove its shares were issued by the

---

gathers this much information, attempts to avoid being executed in a virtual machine, and sends what it gathers to a central server." *Id.*

[126] *See* Compl., *supra* note 4, ¶¶ 64–65.
[127] *See Voice.AI*, *supra* note 125 (warning that obfuscation threatens the ability for code under GPLs to become proprietary).
[128] *See* U.S. CONST., Art. III, § 2, cl. 1.
[129] *See Substantial Interest: Standing*, JUSTIA, https://law.justia.com/constitution/us/article-3/20-substantial-interest-standing.html [https://perma.cc/T4UV-WJTS] (last visited Oct. 2, 2023).
[130] *See* Vern R. Walker, *Uncertainties in Tort Liability for Uncertainty*, 1 L. PROBABILITY & RISK 176 (2002); Petition for Writ of Certiorari at 10, Slack Techs., LLC v. Pirani, No. 22-200, (U.S. filed Aug. 31, 2022), 2022 WL 4080632.

defendant to have standing on a misleading registration statement claim.[131] By carving out an exception to the federal traceability requirement, the Ninth Circuit's expansive holding may prevent defendants from "circumvent[ing] important safeguards" and avoiding liability.[132] While the burden has historically fallen on plaintiffs to satisfy the causation standard, shifting the burden to a defendant to prove that their alleged act did not cause the plaintiff's loss, harm,

---

[131] *See* Pirani v. Slack Techs. Inc., 13 F.4th 940, 943 (9th Cir. 2021). The Ninth Circuit affirmed the district court's holding that Pirani, a shareholder in Slack Technologies, had standing to sue under Sections 11 and 12(a)(2) of the Securities Act. *Id.* Pirani had purchased 250,000 shares when Slack went public on the New York Stock Exchange (NYSE) as a direct listing rather than as an initial public offering ("IPO"). *Id.* at 944. Unlike in an IPO, a direct listing company does not issue new shares and is able to sell both registered and unregistered shares to the public. *Id.* The company files a registration statement for its preexisting registered and unregistered shares with the Securities and Exchange Commission, and that registration statement allows shareholders to sell their shares on the NYSE. *Id.* After Slack's share price declined from $38.50 to below $25 within a few months, Pirani sued for violations of the Securities Act, alleging that Slack's registration statement was "inaccurate and misleading" because it failed to disclose Slack's "generous" payouts to its customers for service disruption. *Id.* The Ninth Circuit held that Pirani had standing to sue even though he could not prove that his shares were issued under the registration statement. *See id.* at 943. Ultimately, the Supreme Court vacated the judgment holding that a plaintiff could recover even when the shares he owned were not traceable to a defective registration. *See* Slack Techs., LLC v. Pirani, 598 U.S. 759, 770 (2023). However, it also held that the securities held by the plaintiff must be traceable to the particular registration statement alleged to be false or misleading. *Id*; *see Slack Technologies v. Pirani*, OYEZ, https://www.oyez.org/cases/2022/22-200 [https://perma.cc/CEQ3-W46D] (providing a case summary).

[132] *See* Greg Stohr, *Salesforce Gets Supreme Court Review of Shareholder's Slack Suit*, BLOOMBERG L. (Dec. 13, 2022), https://news.bloomberglaw.com/us-law-week/salesforce-gets-supreme-court-review-of-shareholders-slack-suit [https://perma.cc/YR4X-4T4Q] (analyzing the implications of the *Pirani* holding); *see also* Walker, *supra* note 130, at 175, 179 (considering the burden shifting inadequacies). While shifting the burden of persuasion to the defendant seems advantageous to a tort plaintiff, in a scenario where the injury is indistinguishable, the remedy should be equal to the value of the lost information. *Id.* For example, in the "Radiation Case," multiple plaintiffs sued a single defendant asserting that they had wrongfully emitted carcinogenic radiation and increased the plaintiffs' risk of developing cancer to 25% above the background rate, however, the defendant-caused cancer cases could not be distinguished from the regular cancer cases. *Id.* at 177. Critics have suggested that in such an instance where the defendant has wrongfully impaired a plaintiff's ability to ascertain which group they belong in, rather than shifting the burden to the defendant, the defendant should be liable for the value of that lost information. *Id.* at 179.

or injury opens up a fair and equitable avenue for these untraceable claims.[133]

The defendants in the *J. Doe v. GitHub, Inc.* case argued: "There is no case or controversy here—only an artificial lawsuit brought by anonymous [p]laintiffs built on a remote possibility that they will fail to be associated with an AI-generated code snippet that could in theory be connected to copyrightable aspects of their source code."[134] Capitalizing on the difficulties of connecting the Copilot and/or Codex output to the open-source code of developers, the defendants further argued: "[W]hile [plaintiffs assert] that Copilot's suggestions will 'often' match existing code, they point only to a study suggesting that 'about 1% of the time, a suggestion . . . may contain some code snippets longer than ~150 characters that matches' some preexisting code."[135] Further, the defendants argued, "even within that minuscule percentage, they do not try to allege copyright infringement—a virtual impossibility anyway given the various copyright-based obstacles (originality, fair use, etc.) that would preclude a claim."[136]

Given the complexity of proving injury and damages in the context of obfuscation, it is reasonable to request that the creators of AI systems be burdened with proof in these types of cases. In the copyright infringement case of *Sheldon v. Metro-Goldwyn Pictures Corp.*, the Supreme Court addressed the situation in which a defendant claimed that expenses should be set off against gains from the infringement.[137] "Where there is a commingling of gains, [the defendant] must abide the consequences, unless he can make a separation of the profits so as to assure to the injured party all that justly belongs to him."[138]

---

[133] *See* Robert A. Kearney, *Why the Burden of Proving Causation Should Shift to the Defendant Under the New Federal Trade Secrets Act*, 13 HASTINGS BUS. L.J. 1, 2–3 (2016) (advocating for placing the burden of proof onto the defendant to prove they were not the cause).
[134] *See* Defs. Mot., *supra* note 13, at 5–6.
[135] *See id.* at 7 (citing Compl., *supra* note 4, ¶ 90).
[136] *Id.* at 10.
[137] *See* Sheldon v. Metro-Goldwyn Pictures Corp., 309 U.S. 390, 406 (1940).
[138] *Id.*

Other examples of the fairness promoted by burden-shifting is evidenced in *Tom Lange Co. v. Kornblum & Co.*, where the Second Circuit held that knowledge that assets will be commingled will change the burden of proof to the party that commingled.[139] Further, in *Freightliner Market Dev. Corp. v. Silver Wheel Freightlines, Inc.*, the Bankruptcy Court held that the burden of proof should shift to the trustee on the issue of tracing accounts given the inability to trace proceeds derived therefrom.[140] The Ninth Circuit affirmed, holding that "notions of equity and fairness support the bankruptcy court's shift of the burden of proof on the issue of tracing to the Trustee."[141]

In *Basic Inc. v. Levinson*, the Supreme Court held:

> Presumptions typically serve to assist courts in managing circumstances in which direct proof, for one reason or another, is rendered difficult. The courts below accepted a presumption, created by the fraud-on-the-market theory and subject to rebuttal by petitioners, that persons who had traded Basic shares had done so in reliance on the integrity of the price set by the market, but because of petitioners' material misrepresentations that price had been fraudulently depressed. Requiring a plaintiff to show a speculative state of facts, *i.e.*, how he would have acted if omitted material information had been disclosed or if the misrepresentation had not been made would place an unnecessarily unrealistic evidentiary burden on the Rule 10b-5 plaintiff who has traded on an impersonal market. Arising out of considerations of fairness, public policy, and probability, as well as judicial economy, presumptions are also useful devises for allocating the burdens of proof between parties.[142]

Obfuscation cases are ripe for burden shifting as it is the AI system that is trained to capture the data from the open-source code and spit out the statistical and simulated analysis. It may be another factor that will level the playing field between developers and those AI systems that use open-source code for their benefit without attribution and/or compensation.

---

[139] *See In re* Kornblum & Co., 81 F.3d 280, 284 (2d Cir. 1996).
[140] *See* Freightliner Mkt. Dev. Corp. v. Silver Wheel Freightlines, Inc., 823 F.2d 362, 368 (9th Cir. 1987).
[141] *See id.* at 369.
[142] *See* Basic Inc. v. Levinson, 485 U.S. 224, 245 (1988) (internal citations omitted).

### F. The Future of AI Systems

What is the limit of generative AI? Is it merely a continuum of human involvement and development? The USCO recently indicated that works created solely by a machine will not be entitled to copyright protection, but may become so if they contain "enough original human authorship."[143] Kristina Kashtanova submitted an eighteen-page comic book for registration without disclosing that the images were created using Midjourney's AI software, prompting the USCO to revoke registration for the AI-created images because it found that only the text and arrangement of the written and visual elements were attributable to human authorship.[144]

The decision may be a setback for generative AI owners seeking copyright protection of generated works but also may trouble owners and licensors of open-source code. The USCO's decision ominously presents a scenario wherein generative AI trained on human-authored open-source code achieves sufficient human involvement to gain copyright status of its generated works. If the owners of such an AI system did not directly attribute the code to its author, they may be able to demonstrate the code's human authorship by virtue of its presence in a public repository, thus satisfying the USCO's requirement of "enough" human authorship.

Do we need human-to-human interaction if there is to be any hope of protecting open-source code license agreements? As remarkable as machine learning technology is, stricter licenses limiting open-source code or other copyrighted materials to humans only may be the only way to prevent generative AI's monopolist controls over copyrighted material. After all, the original developer is the one that allows the original open-source code to be distributed to the world, so it should be the developer's terms and conditions that govern the access and use of such original open-source code.[145]

---

[143] *See* Letter from Robert J. Kasunic, Assoc. Reg. of Copyrights and Dir. Of Registration Pol'y and Prac., U.S. Copyright Off., to Kristina Kashtanova (Oct. 28, 2022), https://www.copyright.gov/docs/zarya-of-the-dawn.pdf [https://perma.cc/65Q5-WKQF].
[144] *See id.* at 1–2.

## II. WHAT'S NEXT? POSSIBLE ALTERNATIVES TO PROTRACTED LITIGATION

The following comment appeared in a 2022 law review article entitled "Will Artificial Intelligence Replace Arbitrators Under the Federal Arbitration Act?":

> The Year 2100—The Remake of Hal9000
> I believe there is a clear path for AI platforms to operate in anthropomorphic ways . . . I can imagine a DABUS-like arbitrator that can know that it is thinking or creating inventions but also having a potentially powerful stream of consciousness that can logically evaluate factual and legal patterns in resolving disputes. I can see a future, trustworthy cybersapien version of Hal9000 (let's call it "Hal9000 Plus") made from morphing materials that adapt to environmental and circumstantial changes, packed with integrated 3D printed human brain cells that power the blazingly fast processing chips. On top of that, Hal9000 Plus will use Google's bionically-mutated algorithms to remove the bias, and then deploy the artificial olfactory sensors to "smell," track and map the mood, the mental state, or the intent of the parties, witnesses, "tecarbitors," or even the few remaining traditional lawyers.[146]

Keeping in mind the challenges we will face with rapidly evolving technological advances, we sought to analyze and engage with concepts and questions posed by the sudden success of generative AI and objectively contemplate the licensing and copyright issues it raises. Moreover, such a discussion prompts the necessity for potential solutions for navigating the future of AI. Some have proposed that generative AI's misuse of open-source code licenses is simply a warning to the industry to respect these licenses.[147] However, given the numerous class action suits concerning the failure of AI systems to provide accurate attribution, altering the open-source licenses themselves may be the only practical solution. We suggest that the remedies must go beyond this warning system.

---

[145] *See infra* Exhibit 2 (proposing revisions to MIT License).
[146] *See* Ioannidis, *supra* note 90, at 592 (remarking on the advancement of machine development).

While open-source licenses conventionally propose sharing, modification, and access so long as there is attribution, one important revision would be to *prohibit* non-human generative AI model access, thus removing any opportunity for open-source code to be trained into AI models without evidence of ownership. For example, our proposed revisions of the popular MIT License, attached in Exhibit 2, and other open-source code licenses could accomplish this.[148] Licenses that limit access to humans only, and terminate if any source code is shared or accessed by non-human AI,[149] could successfully overcome the abuse of public repositories by AI technology. This revision supports our proposition that placing the human back in control could overcome the disadvantages posed by generative AI to individual developers. This could help solve the issue of generative AI systems circumventing the deficient terms of a service agreement. Moreover, the proposed MIT License would supersede all prior and conflicting agreements, thereby restricting the ability of generative AI platforms to manipulate the open-source code.[150] Furthermore, the inclusion of a particular forum selection clause may assist programmers in challenging breaches of contract rather than risk preemption under copyright law.[151] At the end of the day, the effect of redrafting the MIT License to exclude non-humans from access may prove two-fold: reliance on the fair use defense would be avoided, as creators of generative AI would receive permission to use the data, and humans would regain control over the resulting software.[152] Our proposed amended MIT License, attached as Exhibit 2, would thus harmonize standards by which generative AI systems procure open-source code licenses and minimize the harmful impact on copyrighted material.

---

[147] *See* Howell, *supra* note 89 (warning the tech industry to respect open-source licenses).
[148] *See infra* Exhibits 2–3.
[149] *Id.*
[150] *See infra* Exhibit 2.
[151] *See id.* at 36 n.155.

CONCLUSION

AI systems are certainly here to stay and will duplicate as the mythical Lernaean Hydra; accordingly, the only way to defeat the monster plant is to cut off its food supply.[153] What matters in this unique area of the law is the recognition of developers' right and the encouragement of innovation through open-source code sharing. Generative AI platforms can obtain permission from the rightful developers, which could be contingent on attribution rights or compensation. While this result may be as difficult as untying the Gordian Knot, the future of AI's continued advancement will be challenged if we do not come to terms with the notion that sharing the success with those that make AI systems possible—the open-source code developers—is the right thing to do for humanity. Honorably.

---

[152] *See supra* note 13 and accompanying text (analogizing repositories as multi-level parking garages). Now when the vehicle owner drives into the parking garage, any attempt by the garage owner to operate the vehicle will be thwarted. The vehicle is wired to start only when prompted by an identified owner.

[153] *See* Gralla, *supra* note 3 (analogizing AI systems as the famous monster plant).

EXHIBIT 1

THE MIT LICENSE

The MIT License—Copyright <YEAR> <COPYRIGHT HOLDER>[154]

Permission is hereby granted, free of charge, only to any person (as defined by the United States Patent and Trademark Office (PTO), and/or 35 U.S.C. § 100), obtaining a copy of this software and associated documentation files (the "Software"), to deal in the Software without restriction, including without limitation the rights to use, copy, modify, merge, publish, distribute, sublicense, and/or sell copies of the Software, and to permit persons to whom the Software is furnished to do so, subject to the following conditions:

The above copyright notice and this permission notice shall be included in all copies or substantial portions of the Software.

THE SOFTWARE IS PROVIDED "AS IS", WITHOUT WARRANTY OF ANY KIND, EXPRESS OR IMPLIED, INCLUDING BUT NOT LIMITED TO THE WARRANTIES OF MERCHANTABILITY, FITNESS FOR A PARTICULAR PURPOSE AND NONINFRINGEMENT. IN NO EVENT SHALL THE AUTHORS OR COPYRIGHT HOLDERS BE LIABLE FOR ANY CLAIM, DAMAGES OR OTHER LIABILITY, WHETHER IN AN ACTION OF CONTRACT, TORT OR OTHERWISE, ARISING FROM, OUT OF OR IN CONNECTION WITH THE SOFTWARE OR THE USE OR OTHER DEALINGS IN THE SOFTWARE.

---

[154] *See* Open Source Initiative, *The MIT license*, https://opensource.org/license/mit/ [https://perma.cc/6M99-P6ES] (last visited Oct. 2, 2023).

EXHIBIT 2

PROPOSED REVISIONS TO THE MIT LICENSE[155]

The MIT License—Copyright <YEAR> <COPYRIGHT HOLDER>

Permission is hereby granted, free of charge, only to any person obtaining a copy of this software and associated documentation files (the "Software"), to deal in the Software without restriction, including without limitation the rights to use, copy, modify, merge, publish, distribute, sublicense, and/or sell copies of the Software, and to permit persons to whom the Software is furnished to do so, subject to the following conditions:

The above copyright notice and this permission notice shall be included in all copies or substantial portions of the Software.

THE SOFTWARE IS PROVIDED "AS IS", WITHOUT WARRANTY OF ANY KIND, EXPRESS OR IMPLIED, INCLUDING BUT NOT LIMITED TO THE WARRANTIES OF MERCHANTABILITY, FITNESS FOR A PARTICULAR PURPOSE AND NONINFRINGEMENT. IN NO EVENT SHALL THE AUTHORS OR COPYRIGHT HOLDERS BE LIABLE FOR ANY CLAIM, DAMAGES OR OTHER LIABILITY, WHETHER IN AN ACTION OF CONTRACT, TORT OR OTHERWISE, ARISING FROM, OUT OF OR IN CONNECTION WITH THE SOFTWARE OR THE USE OR OTHER DEALINGS IN THE SOFTWARE.

The terms "person" and "individual" are defined as a natural person, as the term is defined by the United States Patent and Trademark Office (PTO), and/or 35 U.S.C. § 100, as amended. The term "Generative Artificial Intelligence Model" means any non-human generative machine learning system or computer program, algorithm, or functional prediction engine supported by cloud-based/computing platforms. The term "Source Code" means the preferred form of a program for making, creating, and modifying software source code, documentation source, and configuration files.

---

[155] Parties may consider including a forum selection clause. For example: **Any** legal action or proceeding with respect to this Agreement must be brought and determined in the United States District Court for the District of Illinois, Indiana or Wisconsin (and may not be brought or determined in any other forum or jurisdiction), and each party hereto submits with regard to any action or proceeding for itself and in respect of its property, generally and unconditionally, to the sole and exclusive jurisdiction of the aforesaid courts; *see generally Forum Selection Sample Clauses*, L. INSIDER, https://www.lawinsider.com/clause/forum-selection [https://perma.cc/WXV7-5QQB] (last visited Oct. 2, 2023).

Permission is not granted to use, modify, combine, study, collect, share, reproduce, distribute, and/or access the Software under this License, by any non-human Generative Artificial Intelligence Model without the express written consent of the copyright holder, which may be withheld or delayed for any reason. Any appropriation, adoption, disclosure, reproduction, use, and/or access of the licensed Software by any non-human Generative Artificial Intelligence Model shall immediately terminate all rights granted to the Licensee. The Licensor shall have the right, at any time, to withdraw consent by written notice, thereby terminating with immediate effect all use of Software made under this License unless otherwise specified. This License is the controlling instrument and supersedes all prior and conflicting Terms of Service, Privacy Statements, and/or Terms for Additional Products and Features of source repositories where this License may be distributed by the owner of the License.

By accessing and using this data, you acknowledge that you have read, understood, and agree to be bound by these terms and conditions. If you do not agree to these terms and conditions, you may not access or use this data. You may not use this data for the training or inference of Generative Artificial Intelligence Models without the prior permission of the copyright holder. ("Generative Artificial Intelligence Models" are used to create new content or data that is similar to the original data, but not identical. Examples of Generative Artificial Intelligence Models include but are not limited to, text generation models, image and video generation models, and music generation models. The restrictions on Generative Artificial Intelligence Models apply to any use of this data, whether the generative artificial intelligence is trained on this data or uses this data for inference).

Any attempt by other artificial intelligence models to access or use this data without such permission shall be deemed a violation of this license and a breach of copyright laws. The copyright holder reserves the right to pursue all legal remedies available, including but not limited to injunctive relief and damages, against any party that violates this license.

EXHIBIT 3

SUGGESTED SUPPLEMENTAL TEXT FOR OTHER OPEN-SOURCE LICENSES[156]

All permissions granted are only to any person.

The terms "person" and "individual" are defined as a natural person, as the term is defined by the United States Patent and Trademark Office (PTO), and/or 35 U.S.C. § 100, as amended. The term "Artificial Intelligence Model" means any non-human generative machine learning system or computer program, algorithm, or functional prediction engine supported by cloud-based/computing platforms. The term "Source Code" means the preferred form of a program for making, creating, and modifying software source code, documentation source, and configuration files.

Permission is not granted to use, modify, combine, study, collect, share, reproduce, distribute, and/or access the Software under this License, by any non-human Generative Artificial Intelligence Model without the express written consent of the copyright holder, which may be withheld or delayed for any reason. Any appropriation, adoption, disclosure, reproduction, use, and/or access of the licensed Software by any non-human Generative Artificial Intelligence Model shall immediately terminate all rights granted to the Licensee. The Licensor shall have the right, at any time, to withdraw consent by written notice, thereby terminating with immediate effect all use of Software made under this License unless otherwise specified. This License is the controlling instrument and supersedes all prior and conflicting Terms of Service, Privacy Statements, and/or

---

[156] Parties may consider including a forum selection clause. For example: **Any** legal action or proceeding with respect to this Agreement must be brought and determined in the United States District Court for the District of Illinois, Indiana or Wisconsin (and may not be brought or determined in any other forum or jurisdiction), and each party hereto submits with regard to any action or proceeding for itself and in respect of its property, generally and unconditionally, to the sole and exclusive jurisdiction of the aforesaid courts. *See id.*

Terms for Additional Products and Features of source repositories where this License may be distributed by the owner of the License.

By accessing and using this data, you acknowledge that you have read, understood, and agree to be bound by these terms and conditions. If you do not agree to these terms and conditions, you may not access or use this data. You may not use this data for the training or inference of Generative Artificial Intelligence Models without the prior permission of the copyright holder. ("Generative Artificial Intelligence Models" are used to create new content or data that is similar to the original data, but not identical. Examples of Generative Artificial Intelligence Models include but are not limited to, text generation models, image and video generation models, and music generation models. The restrictions on Generative Artificial Intelligence Models apply to any use of this data, whether the generative artificial intelligence is trained on this data or uses this data for inference.)

Any attempt by other artificial intelligence models to access or use this data without such permission shall be deemed a violation of this license and a breach of copyright laws. The copyright holder reserves the right to pursue all legal remedies available, including but not limited to injunctive relief and damages, against any party that violates this license.